\documentclass[
pre,twocolumn,notitlepage,longbibliography,amsmath,amssymb,floats,superscriptaddress,nofootinbib,10pt]{revtex4-2}
\usepackage{graphicx,color}
\usepackage{amsmath}
\usepackage{amsfonts, amsbsy}
\usepackage{amssymb}
\usepackage{eqnarray}
\usepackage{bm}
\usepackage{blkarray}
\usepackage{mathtools}
\usepackage{psfrag}
\usepackage{multirow}
\usepackage{textcomp}
\usepackage{units}
\usepackage{lipsum}
\usepackage[title]{appendix}
\usepackage{soul}
\usepackage{titlesec}
\usepackage{times}
\usepackage{enumitem}
\usepackage{comment}
\usepackage{braket}
\usepackage[sort&compress]{natbib}
\usepackage{array}

\usepackage[breaklinks, colorlinks = true, linkcolor = blue, urlcolor = blue, citecolor = magenta, anchorcolor = blue]{hyperref}
\usepackage{float}
\usepackage{orcidlink}

\usepackage{xcolor}

\DeclareMathOperator{\csch}{csch}

\begin{document}
\title{Curved Odd Elasticity}

\author{Yuan Zhou\orcidlink{0000-0002-2481-205X}}
\email{y.zhou@uva.nl}
\affiliation{Institute of Physics, University of Amsterdam,
Science Park 904, 1098 XH, Amsterdam, The Netherlands}

\author{Lazaros Tsaloukidis\orcidlink{0000-0003-1292-1037
}}
\email{ltsalouk@pks.mpg.de}
\affiliation{Max Planck Institute for the Physics of Complex Systems, N\"othnitzer Str. 38, 01187, Dresden, Germany}
\affiliation{W\"urzburg-Dresden Cluster of Excellence ct.qmat, 01187, Dresden, Germany}

\author{Jack Binysh\orcidlink{0000-0002-4880-0942}}
\email{j.a.c.binysh@uva.nl}
\affiliation{Institute of Physics, University of Amsterdam,
Science Park 904, 1098 XH, Amsterdam, The Netherlands}

\author{Yuchao Chen}
\email{changyu@mit.edu}
\affiliation{Department of Physics, Massachusetts Institute of Technology, Cambridge, MA, USA}

\author{Nikta Fakhri\orcidlink{0000-0003-1261-7465}}
\email{fakhri@mit.edu}
\affiliation{Department of Physics, Massachusetts Institute of Technology, Cambridge, MA, USA}

\author{Corentin Coulais\orcidlink{0000-0002-3174-5836}}
\email{coulais@uva.nl}
\affiliation{Institute of Physics, University of Amsterdam,
Science Park 904, 1098 XH, Amsterdam, The Netherlands}

\author{Piotr Sur\'owka\orcidlink{0000-0003-2204-9422}}
\email{piotr.surowka@pwr.edu.pl}
\affiliation{Institute of Theoretical Physics, Wroc\l{}aw University of Science and Technology, 50-370 Wroc\l{}aw, Poland}

\thanks{LT and YZ contributed equally to this work.}
\date{\today}

\begin{abstract}
Living materials such as membranes, cytoskeletal assemblies, cell collectives and tissues can often be described as active solids---materials that are energized from within, with elastic response about a well defined reference configuration. These materials often live in complex and curved manifolds, yet most descriptions of active solids are flat. Here, we explore the interplay between curvature and non-reciprocal elasticity via a covariant effective theory on curved manifolds in combination with numerical simulations. We find that curvature spatially patterns activity, gaps the spectrum, modifies exceptional points and introduces non-Hermitian defect modes. Together these results establish a foundation for hydrodynamic and rheological models on curved manifolds, with direct implications for living matter and active metamaterials.
\end{abstract}

\maketitle

 {\it Introduction}$-$Many biological and engineered structures---vesicles, cell monolayers, metamaterials---are active solids: materials that are energized from within,  with elastic response about a well-defined reference state. Pumping energy into elastic materials brings a range of distinctive phenomena: waves propagate in overdamped media~\cite{tan2022Odd,xuAutonomousWavesGlobal2023,baconnierSelfaligningPolarActive2025}, modes condense~\cite{liuViscoelasticControlSpatiotemporal2021,baconnierSelectiveCollectiveActuation2022a,guEmergenceCollectiveOscillations2025}, topological defects nucleate shape change~\cite{maroudas-sacksTopologicalDefectsNematic2021,braunsActiveSolidsTopological2025}, and solids locomote by themselves~\cite{Veenstra2025Adaptive}.
 The dominant paradigm in active matter has been fluid-like response, and active stresses that carry an anisotropic order---either polar or nematic~\cite{shankarTopologicalActiveMatter2022}. However, recent experiments on e.g. colloids~\cite{bililign2022Motile}, metamaterials~\cite{chenRealizationActiveMetamaterials2021,Veenstra2025Adaptive}, spinning embryos~\cite{tan2022Odd}, or cellular monolayers~\cite{chenChiralityScalesTissue2025} emphasize chiral active solids as a distinct class of active matter, which features stress propagation without flow, and persistent oscillations. These systems call for complementary elastic descriptions. 
\begin{figure}[t!]
\includegraphics[width=0.95\linewidth]{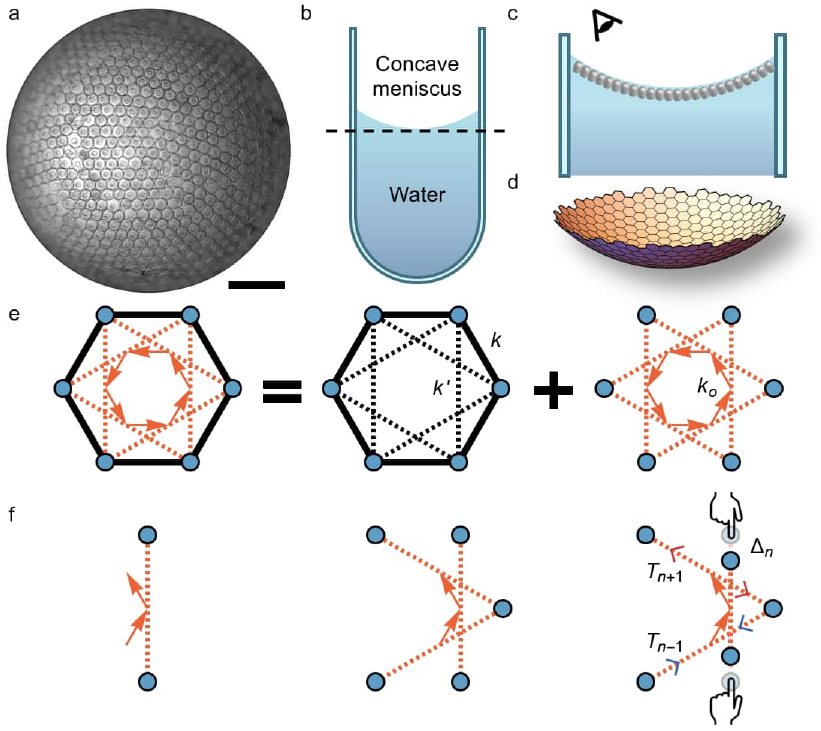}
\caption{{\bf Curved odd elasticity.}
    (a) A living chiral crystal composed of starfish embryos assembled on a curved fluid interface, illustrating an experimental realization of active crystalline order on a curved manifold. Scale bar, 1 mm.
   (b) Curved meniscus. 
   (c) Side cross-sectional schematic showing curved living chiral crystal (visible as top view in (a)) formed below the meniscus.
   (d) Tessellating active plaquettes on curved surfaces provides a minimal model to study how curvature and topology modify non-reciprocal mechanical response. 
   (e) Unit cell of a honeycomb lattice, consisting of passive nearest- and next-nearest-neighbor bonds with spring stiffness $k$ and $k'$, together with nonreciprocal bonds obeying the tension-length relationship $T_n=k_o(\Delta_{n+1}- \Delta_{n-1})$. Here $T_n$ is the tension in bond $n$, and $\Delta_{n+1}$, $ \Delta_{n-1}$ are the length changes in the neighboring bonds $n+1$ and $n-1$ within a hexagonal plaquette.
   (f) An example deformation, in which bond $n$ is compressed with length change $\Delta_n < 0$, causing the neighboring bonds to experience tension $T_{n+1} = -k_o \Delta_n$ and compression $T_{n-1} = k_o \Delta_n$.
   }
   \label{fig:manifold}
\end{figure}

In this context, the framework of odd elasticity captures chiral response in a concise and compelling fashion  \cite{Scheibner2020,Fruchart2023OddViscosityElasticity,Lakes2025Colloquium}.
But odd elasticity is a flat-space theory, and active materials are often curved, especially in biology~[Fig. \ref{fig:manifold}(a--c)]. This curvature matters: curvature-induced distortions can cause localized vibrational modes not present in flat space~\cite{Shankar2017}, and topologically enforce defects that act as stress foci~\cite{keberTopologyDynamicsActive2014, sawTopologicalDefectsEpithelia2017, maroudas-sacksTopologicalDefectsNematic2021}. Therefore a central question is: How do odd active crystals interplay with geometry---specifically, how do curvature and topology modify, enable, or suppress the non-reciprocal mechanical responses and collective dynamics which are characteristic of odd elasticity?

Here we develop a covariant low-energy theory of odd elasticity on curved manifolds, complemented by microscopic simulations of curved non-reciprocal lattices. Our covariant symmetry-based framework bridges continuum theories of ordered and defective media on curved surfaces 
\cite{efrati_elastic_2009,moshe_elastic_2015,li_elasticity_2019, Shin2008,Napoli2016,Zhang2016,Shankar2017,Henkes2018,He2022,Napoli2021,Wittmann2021}
with the differential-geometric theory of shells and thin elastic bodies~\cite{ciarlet,kamien_geometry_2002,Bowick2009}, while proposing a covariant extension of elasticity theory to non-reciprocal media. We find that curvature gaps complex-frequency modes at long wavelengths, whilst inducing highly localized oscillations around topologically enforced defects. At open boundaries we discover long-wavelength Rayleigh modes that destabilize bulk odd elastic materials from the edge inwards. In a generic curved odd medium, all of these classes of mode exist simultaneously. Our results guide the search for non-reciprocal moduli in complex biomaterials and offer design principles for active metamaterials embedded in our three-dimensional world. 

\begin{figure*}[t]
    \includegraphics[width= \linewidth]{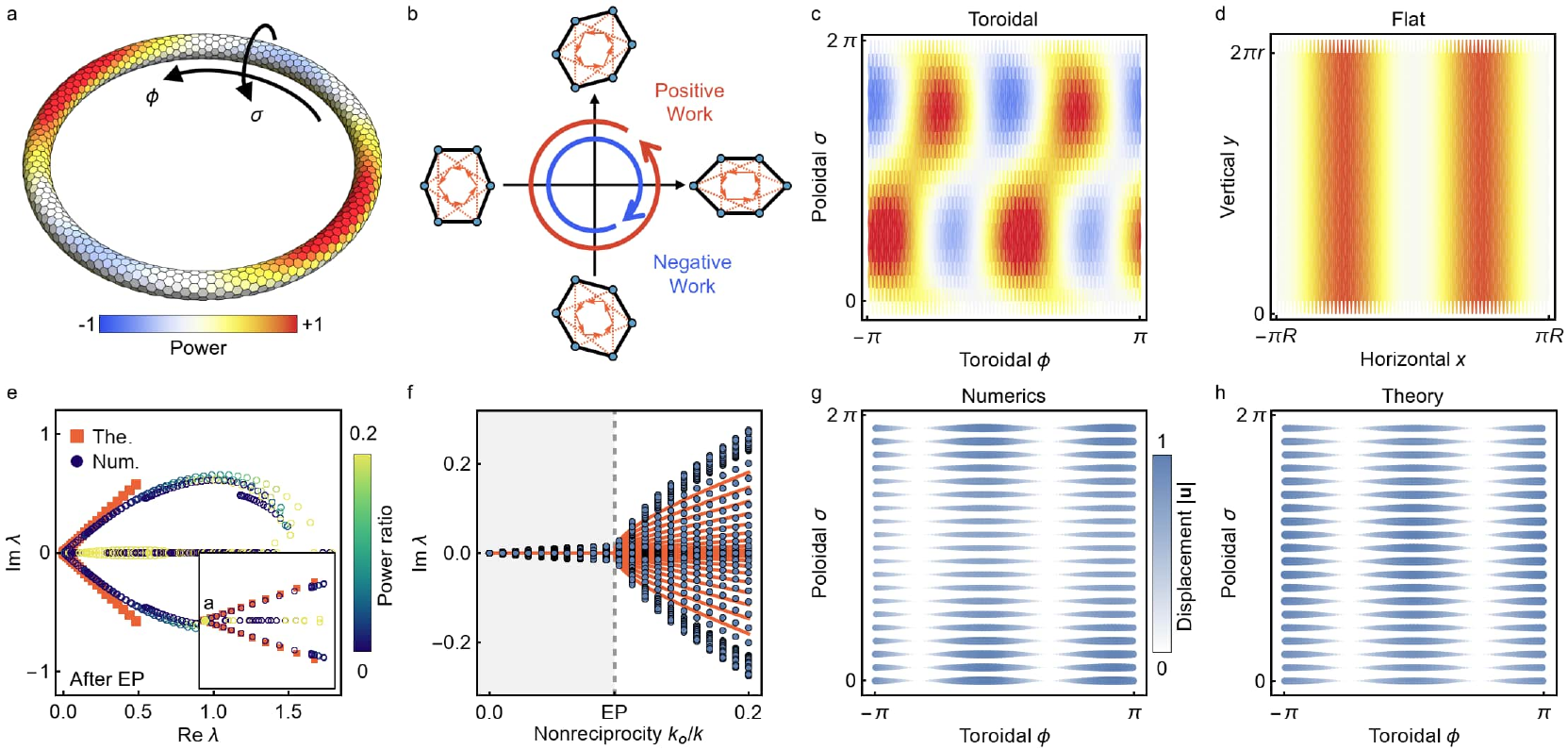}
    \caption{ {\bf Covariant odd elasticity captures the vibrational spectrum of curved non-reciprocal surfaces.} 
    {
    (a, b) An example high-activity bulk mode of a non-reciprocal toroidal lattice. Curvature textures the spatial structure of the modes, creating regions of local power reversal within the lattice in which some plaquettes locally consume energy.  
    (c, d) Regions of power reversal only occur on curved surfaces, not in flat space.
    (e) The vibrational eigenspectrum $\lambda$ of the lattice is non-Hermitian. Colouring each eigenmode by the power ratio $\log(|1+p|/|1-p|)$, where $p=|P^{-}|/|P^{+}|$ quantifies the extent of power reversal, shows that spatial heterogeneity is generic across the spectrum. Inset: The filled marker corresponds to the lowest toroidal mode ($m=1, n=0$) in panels a and c.
    (f) Our covariant formalism Eq.~\eqref{eq:EOM} captures the long-wavelength vibrations of the lattice, and correctly predicts the critical $k_o^*/k=\sqrt{3}(1+6k'/k)/18$ at which the lattice undergoes a bulk exceptional transition and long-wavelength modes become complex.
    (g, h) Our covariant theory captures not only the eigenvalues $\lambda$, but also the spatial structure of the vibrations themselves. Color shows $|\mathbf{u}|$ along the toroidal and poloidal directions for an eigenmode with matched eigenvalue $\lambda$ between our numerics (g) and theory (h). In this study, we set $k'/k = 10^{-4}$, with constraint springs $k_c$ set to $k_c/k=10^4$.
    }}
    \label{fig:torus}
\end{figure*}

\emph{Odd crystallography on curved substrates}$-$
Active crystals made of sea-star embryos living at air-water interfaces have been shown to display odd elasticity \cite{tan2022Odd,chao2026Selective}. These crystals can also live on curved air-water interfaces---menisci [Fig.~\ref{fig:manifold}(a--c)].
Inspired by these systems and by vertex models of curved active solids \cite{sussmanInterplayCurvatureRigidity2020, maroudas-sacksTopologicalDefectsNematic2021}, we construct a microscopic model of active agents with chiral symmetry breaking on a curved substrate. We focus on a minimal model that accounts for self-oscillations while conserving angular momentum: a honeycomb lattice with neighbor interactions that break microscopic reciprocity
[Fig.~\ref{fig:manifold}(d)]. 
We endow each edge of our honeycomb with a minimal non-reciprocal tension-length relationship: $T_n =k_o (\Delta_{n+1}-\Delta_{n-1})$, where $T_n$ is the tension in the $n^{th}$ edge, and $\Delta_{n-1}$ and $\Delta_{n+1}$ denote the length changes of neighboring edges measured in the counterclockwise direction [Fig.~\ref{fig:manifold}(e, f)]. The passive rigidity of the lattice is ensured by nearest-neighbor springs of stiffness $k$ and next-nearest-neighbor of stiffness $k'$. 
Many biophysical systems exhibit collective motions confined to a curved substrate \cite{brandstatter2023Curvature}. We therefore add strong geometric constraints that confine nodal displacements to the tangent plane of the surface. In flat space, the long-wavelength response of this model has recently been shown to possess odd elastic moduli~\cite{binysh2025More}. How does curvature affect this coarse-grained response?

Curved surfaces bring two ingredients: geometric distortions through a spatially varying metric, and topological constraints through the Euler characteristic. To disentangle these effects we first focus on geometry alone and consider a surface with Euler number zero: a torus [Fig.~\ref{fig:torus}(a)]. Diagonalizing the non-Hermitian dynamical matrix $\mathbf{D}$ of our toroidal lattice, we find complex  eigenvalues $\lambda$ with eigenvectors $\bf u$ [SM \S V]~\cite{lubensky2015Phonons,binysh2025More}. We interpret the imaginary component of $\lambda$ assuming an overdamped dynamics, as is typically the case in biophysical materials~\cite{tan2022Odd}. In this context, $\mathrm{Im}(\lambda)\neq 0$ corresponds to oscillations. However, inertial dynamics is also possible~\cite{veenstra_wave_2025}, in which case $\mathrm{Im}(\lambda)>0$ implies linear instability.
In flat space, the corresponding eigenmodes would be simply periodic~\cite{Scheibner2020} and these standing wave modes exhibit consistent strain cycles and energy injection. 
By contrast, here we find that curvature gives the eigenmodes spatial texture [Fig.~\ref{fig:torus}(a-c)].

To further quantify this spatial structure, we check the local energy injection of each plaquette $P_i$ by calculating the power of internal nonreciprocal bonds $P=\sum_{\text{plaquette}}P_i \propto \operatorname{Im}(\mathbf{u}^\dagger \mathbf{D} \mathbf{u})$. We find a distribution of power injection and dissipation: some hexagonal units have $P_i<0$ even though the whole system has a net positive power [Fig.~\ref{fig:torus}(c)]. To measure the degree of power reversal in each eigenmode, we sum up the positive $P^+=\sum (P_i>0)$ and negative $P^-=\sum (P_i<0)$ power contributions and introduce the power ratio $p=|P^{-}|/|P^{+}|$. We color each mode according to $\log(|1+p|/|1-p|)$, which has maximum value when the mode has comparable positive and negative power. In contrast to flat systems, we find that many modes indeed exhibit spatial heterogeneity in their power dissipation---a clear experimentally accessible signature of curvature [Fig.~\ref{fig:torus}(e)]. The question now is how to capture such spatial texture theoretically.

 {\it A covariant formulation of odd elasticity}$-$To explain the spatial structure of our eigenmodes,
 we now develop a framework of curved odd elasticity for a general strain metric. Crystallography on curved manifolds \cite{Vitelli2006Crystallography} is most naturally formulated in a covariant, two–metric language \cite{efrati_elastic_2009} in which crystalline order {spontaneously breaks} continuous translations (and rotations). We take a stress–free \emph{reference} metric $\bar g_{ij}$ encoding the undeformed lattice and a \emph{dynamical} metric $g_{ij}$ for the deformed configuration, with symmetric strain $
u_{ij}=\tfrac12\!\left(g_{ij}-\bar g_{ij}\right)
$. Within this framework, stresses $\sigma^{ab}$ relate to strains $u_{cd}$ through the elasticity tensor $\sigma^{ab}= C^{abcd}u_{cd}$, and there is a unique covariant extension of the flat-space odd elastic tensor to curved space:
\begin{equation}
C_{odd}^{abcd}=\frac{K_o}{2}\left(\Bar{g}^{ac}\bar{\epsilon}^{bd}+\Bar{g}^{ad}\bar{\epsilon}^{bc}+\Bar{g}^{bc}\bar{\epsilon}^{ad}+\Bar{g}^{bd}\bar{\epsilon}^{ac}\right),
\vspace{2mm}
\label{eq:OddC}
\end{equation}
where $\bar{\epsilon}_{ij}$ is the Levi-Civita tensor density and $K_o$ is the odd modulus. We include this odd contribution directly into the constitutive law built from conservative elasticity [SM~\S I] and derive the active Navier-Cauchy equations of motion~\cite{Tsaloukidis2024}:
\begin{equation}
\begin{aligned}
 \gamma \frac{\partial u_i}{\partial t}  
 = \overline{\nabla}_i\!\left[(B-\mu)\,\overline{\nabla}_j u^j + K_o \bar{\epsilon}^{jk}\,\overline{\nabla}_j u_k\right] \\
+\overline{\nabla}_j\overline{\nabla}^j\!\left[2\mu\,u_i + K_o \bar{\epsilon}_{i}^{k}\, u_k\right] 
 + K_o \bar{\epsilon}_{i}^{k}\,\overline{\nabla}_k\left(\overline{\nabla}_j u^j\right), 
\end{aligned}
\label{eq:oddNavier}
\end{equation}
where $\overline{\nabla}$ is the covariant derivative associated with $\bar g_{ij}$, defined using the Levi–Civita connection, 
and $B,\mu$ are the bulk and shear moduli. These continuum moduli have simple expressions in terms of the microscopic parameters. For our honeycomb lattices, $K_o=3k_o/2$, $B=(k+6k')/(2\sqrt{3})$ and $\mu=\sqrt{3}k'/2$, see SM~\S IV. 
We illustrate Eq.~\eqref{eq:oddNavier} with an overdamped dynamics $\gamma$, but inertial motion is equally possible~\cite{Veenstra2025Adaptive,veenstra_wave_2025}. Beyond the specific case of odd solids, our covariant two–metric construction provides a minimal and broadly applicable framework for incorporating activity into solid elasticity at the level of constitutive relations (see [SM~\S I]). In this baseline formulation, non-reciprocal or active effects are captured through generalized material coefficients in the elasticity tensor $C^{abcd}$, without presupposing additional internal variables, polarity fields, or microstructural order. This does not preclude such fields---which may be essential in more complex active materials---but establishes a geometrically consistent foundation onto which they can be systematically added. Once the reference metric $\bar g_{ij}$ and the symmetry class of the solid are specified, the resulting active modifications of the stress-strain relation are encoded directly in the generalized elastic moduli.

We now use the Helmholtz-Hodge decomposition to separate the longitudinal and transverse modes of Eq.~\eqref{eq:oddNavier} into two potentials, $\chi$ and $\psi$:
$u_i=\overline{\nabla}_i \chi+\bar{\epsilon}_{ij}\overline{\nabla}^j \psi$. We omit the harmonic contribution in this decomposition, since it is annihilated by the Laplace--Beltrami operator and decouples from the linear odd-elastic dynamics considered here.
The dynamics reduces to two coupled scalar equations:
\begin{equation}
    \gamma \frac{\partial}{\partial t}
    \begin{pmatrix}
        \chi \\ \psi
    \end{pmatrix}
    =\overline{\nabla}^2
    \begin{pmatrix}
        B+\mu & -K_o \\
        K_o & \mu
    \end{pmatrix}
    \begin{pmatrix}
        \chi \\ \psi
    \end{pmatrix}.
    \label{eq:EOM}
\end{equation}
Equation~\eqref{eq:EOM} generalizes odd elasticity to any torsionless manifold endowed with a metric-compatible connection. The flat space Laplacian becomes the Laplace-Beltrami operator $\overline{\nabla}^2$, which encodes any curved substrate. The mixing of the scalar potentials $\chi$ and $\psi$ through $K_o$ generalizes the hybridization of phonons found in flat space~\cite{Scheibner2020}. We now use our theory Eq.~\eqref{eq:EOM} to rationalize the vibrational modes of the non-reciprocal torus.

{\it Odd elastic modes of the torus}$-$ 
In the limit of a slender torus with minor radius $r$ much smaller than major radius $R$, we find an analytical solution for the long wavelength eigenspectrum of Eq.~\eqref{eq:EOM} [SM \S III]: 
\begin{equation}
\lambda =\frac{B+2\mu \pm \Delta}{2}\,
\frac{\alpha^2}{r^2 (\alpha^2-1)}
\left(\frac{m^{2}}{\alpha^2-1}+n^{2}\right),
\label{eq:frequency_torus}
\end{equation}
where $m,n$ label the toroidal wavenumber~\cite{Moon1988}, $\alpha=R/r \gg 1$ is the aspect ratio and $\Delta=(B^2-4K_o^2)^{1/2}$.  
The eigenvalues given by Eq.~\eqref{eq:frequency_torus} capture the long-wavelength modes of our numerics [Fig.~\ref{fig:torus}(e)]. We find an exceptional transition at $\Delta=0$, at which the coupling matrix Eq.~\eqref{eq:EOM} becomes defective and the long-wavelength spectrum becomes complex. In the SM~\S IV we explicitly coarse grain the elastic moduli of our lattice to predict a threshold non-reciprocity for this transition of $k_o^*/k=\sqrt{3}(1+6k'/k)/18$, in good agreement with the threshold observed in our numerics [Fig.~\ref{fig:torus}(f)]. Our theory not only captures the mode structure of the eigenvalues $\lambda$, but also the spatial structure of the eigenvectors $\bf u$. For eigenvalues $\lambda$ matched between theory and numerics, we project our numerical displacement field $\bf u$ onto the torus, and find good agreement with solutions derived from Eq.~\eqref{eq:EOM} [Figs.~\ref{fig:torus}(f, g)]. In conclusion, our covariant theory correctly captures the large scale collective modes of curved odd systems, and provides a theoretical baseline to compare with additional lattice-scale vibrations.

\begin{figure}[t]
    \centering
    \includegraphics[width=0.95 \linewidth]{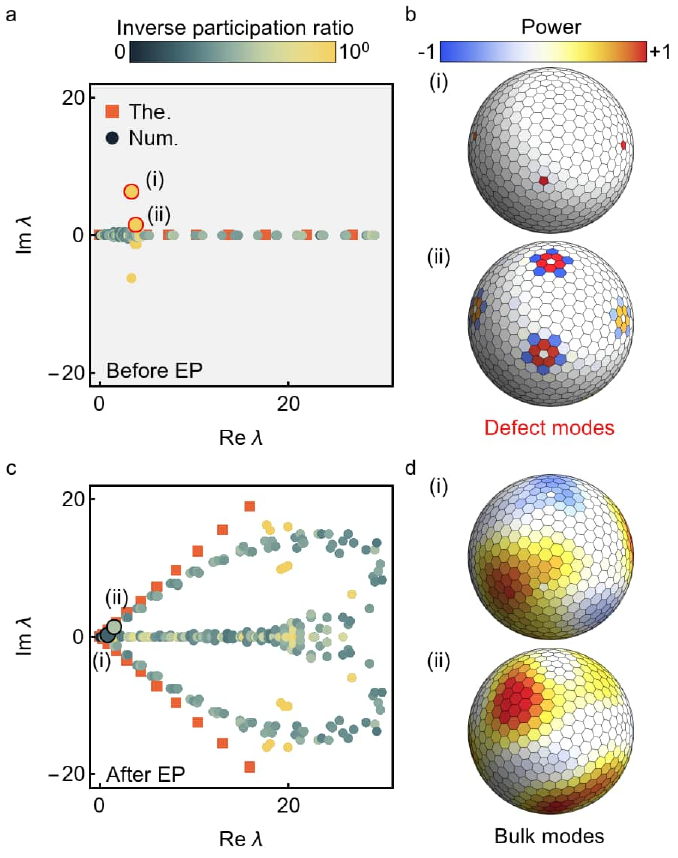}
    \caption{
    {\bf Topological constraints on a sphere enforce active defect-bound modes.} (a, b) Defect-bound modes exhibit highly localized strain fields around topologically enforced disclinations.
    These modes become complex for any nonzero $k_0$, well before the bulk exceptional transition. (c, d) Bulk modes display spatially distributed strain fields and are well-described by our theory Eq.~\eqref{eq:EOM}. This band of continuum modes destabilizes only after the exceptional transition, and merges with the defect-bound modes at large activity. 
    }
    \label{fig:sphere}
\end{figure}
{\it Topological constraints and defects on an odd sphere}$-$ 
Curvature brings not only geometric distortions, but also topological constraints: on surfaces with nonzero Euler characteristic we must find topological defects in our crystallographic lattice. 
To this end we examine the spectrum of spherical polyhedra with twelve pentagons, formed by meshing an icosahedron. 
As in the eigenspectrum of a torus [Fig.~\ref{fig:torus}] we see a long wavelength exceptional transition. However, before this transition we observe a new class of instabilities.

We now find thresholdless modes that oscillate for any nonzero $k_o$ [Fig.~\ref{fig:sphere}(a, b)] \cite{PhysRevLett.127.268001,bililign2022Motile}. Measuring the inverse participation ratio of these modes $ \sum_{\text{nodes}} |u_i|^4$, we find that they are highly localized to lattice defects, and the degeneracy equals the number of pentagons (see SM~\S V, for the degenerate defect modes). Before the exceptional transition, strain concentrates at these defects, which serve as local energy sources and dominate the dynamics. After the exceptional transition, bulk modes also become oscillatory [Fig.~\ref{fig:sphere}(c, d)]. As $k_o$ increases, these bulk bands eventually merge with the isolated defect-bound modes to once again dominate the eigenspectrum.

To capture the spectrum of the odd sphere at the low-frequency long-wavelength limit we now apply the general framework Eq.~\eqref{eq:EOM} to a sphere of radius $r$ [SM \S II]. We find that the eigenmodes are spherical harmonics with spectrum
\begin{equation}
\lambda=\frac{B+2\mu \pm \Delta}{2}\; \frac{l(l+1)}{r^2}.
\label{eq:frequency_sphere}
\end{equation}
Finite curvature opens a finite-size gap $\lambda\sim 1/r^2$, which closes as the space tends to be flat $r\to\infty$, consistent with Goldstone behavior. The analytical prediction also shows a clear level structure 
that agrees well with our numerical results [Fig.~\ref{fig:sphere}(c)]. In summary, curvature simultaneously gaps long-wavelength bulk modes, and introduces highly localized defect-bound states.

\begin{figure}[t!]
    \centering
    \includegraphics[width=0.95 \linewidth]{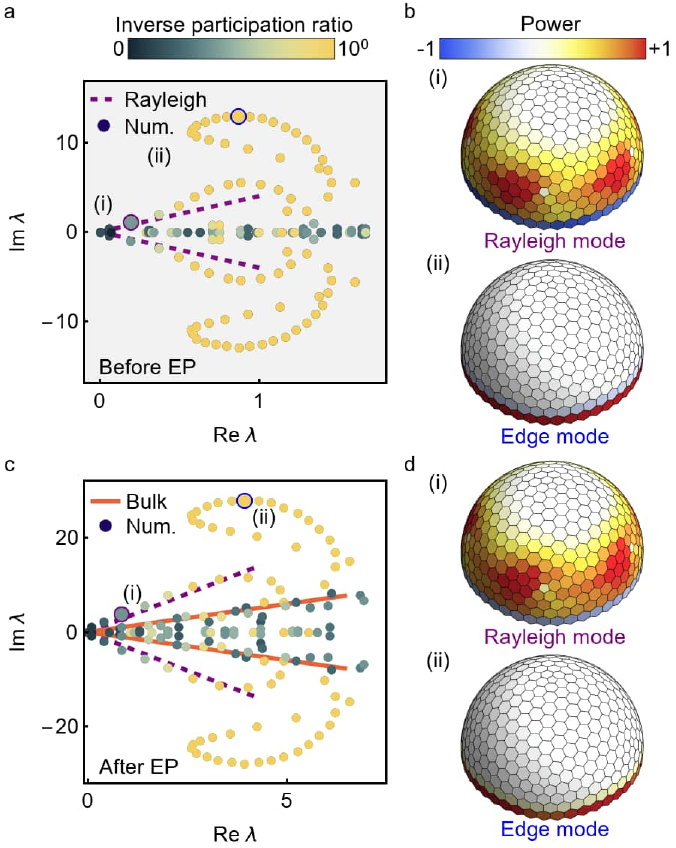}
    \caption{
    {\bf Open boundaries create non-Hermitian Rayleigh modes.}
    (a, b) Cutting a boundary into a spherical cap creates (i) Rayleigh modes with finite penetration depth and (ii) lattice-scale edge modes. Plotting the inverse participation ratio confirms that the low frequency Rayleigh modes are delocalized relative to the edge modes. The purple dashed line denotes our prediction for Rayleigh waves [see SM \S VI, Eq.~(S59)].
    (c, d) In an open cap defect-bound modes, lattice-scale edge modes and Rayleigh waves all co-exist, each becoming complex before the bulk exceptional transition. Rayleigh modes sit above our theoretically predicted bulk modes even beyond the exceptional transition: the edge is always more unstable than the bulk in odd elastic materials.
    The orange line denotes the bulk prediction [see SM \S VI, Eq.~(S60)].
    }
    \label{fig:sphericalcap}
\end{figure}

{\it Edge modes on a spherical cap}$-$ 
We now consider a final class of mode: boundary states, which play an important role in active systems from topological modes \cite{sone2020Exceptional,tang2021Topology} to edge-localized flows \cite{scholz2018Rotating,yang2021Topologically}. We cut an open boundary into our odd sphere to form a cap [Fig.~\ref{fig:sphericalcap}]. Below the bulk exceptional transition we now find two new classes of mode. First are lattice scale edge modes, which have strain cycles localized to the honeycomb plaquettes that form the open boundary itself [Fig.~\ref{fig:sphericalcap}(a, b)]. However, in the long-wavelength regime we also find a new band of surface-bound modes with a finite decay length: These are Rayleigh waves (see SM~\S VI, for Rayleigh waves in flat and curved systems). These Rayleigh modes are thresholdless, like our defect-bound modes [Fig.~\ref{fig:sphere}(a)]: they cause strong boundary oscillations for any nonzero $k_o$, reminiscent of the exceptional edge modes \cite{sone2020Exceptional} and topological edge-gain effect \cite{song2020Symmetric} in non-Hermitian systems. However, unlike our defect-bound modes, even beyond the bulk exceptional transition we find that Rayleigh modes sit at $\mathrm{Im}(\lambda)$ values above our theoretically predicted bulk modes for any $k_o$ [Fig.~\ref{fig:sphericalcap}(c, d)]: the edge always oscillates more strongly than the bulk .
Moreover, these edge modes are robust against variations in the boundary geometry (see SM~\S V, for the robust edge modes).

{\it Discussion}$-$We have built a covariant theory of odd crystals on curved manifolds, complemented by discrete simulations to probe lattice-scale dynamics. 
Taken together, our results suggest a unified physical picture: activity defines how a solid oscillates, while geometry determines where and in what form it oscillates. In flat living chiral crystals, nonreciprocity partitions the dynamics into acoustic and optical families with characteristic work cycles~\cite{chao2026Selective}. 
On curved manifolds, the same odd-elastic principles combine with Gaussian curvature, defects, and boundaries to orchestrate a broader hierarchy of oscillatory modes, ranging from bulk odd phonons to defect-centered micro-oscillators and boundary-localized Rayleigh waves. 
These signatures are experimentally accessible via particle image velocimetry and single-particle tracking methods developed for living chiral crystals---in SM~\S V we give a structural analysis which exhibits defects within the curved crystal shown in Fig.~\ref{fig:manifold}(a).
This geometric control of oscillatory and work-generating cycles offers predictive design principles for biological tissues, synthetic active metamaterials, and living solids embedded in curved environments.

Looking ahead, the same program that begins with the passive covariant theory of elasticity and extends the equations to include active and odd couplings can be further enriched by torque density terms~\cite{Caprini2025OddActive, Lee2025, Banerjee2025Emergent}, Cosserat or micropolar structure~\cite{Surowka2023OddCosserat, Wu2023MicropolarBeam, 
jiang_energy_2025}, and the inclusion of additional reactive moduli~\cite{Christensen2024Perspective}. The continuum theory can also be further generalized to include the out-of-plane motions when substrate softness is comparable to in-plane stiffness~\cite{al-izzi_chiral_2023,fossati_odd_2024,Wolfgram2025OddShells, lai2024Odd, Lai2025SphericalShells}. More broadly, our framework provides a basis for formulating hydrodynamic and rheological theories in a fully covariant manner on curved manifolds. In particular, strains are defined geometrically through intrinsic and induced metrics, while the dynamics is formulated using covariant derivatives and tensorial constitutive relations. Natural extensions include viscoelastic or active chiral media on curved substrates, where curvature and topology can couple to Hall viscosity, odd elasticity, and odd thermo- or viscoelastic transport \cite{
Banerjee2021, 
OstojaStarzewski2024OddThermoelasticity,  
Zhang2024AnisotropicOdd, 
Cohen2025OddDipole,
armas_hydrodynamics_2025}, thereby modifying collective modes, stress propagation, and defect dynamics.


{\it{Acknowledgments}}$-$We thank Akash Bandhoe and Jonas Veenstra for preliminary numerical explorations and insightful discussions. J.B.  acknowledges funding from the European Union’s Horizon research and innovation programme under the Marie Skłodowska-Curie Grant Agreement No. 101106500. C.C. and Y.Z. acknowledge funding from the Netherlands Organization for Scientific Research under grant agreement VI.Vidi.213.131 (C.C. and Y.Z.) and from the European Research Council under grant agreement ERC-CoG  101170693 (C.C.). P.S. was supported in part by the Polish National Science Centre (NCN) Sonata Bis grant 2019/34/E/ST3/00405. L.T. was supported in part by the Deutsche Forschungsgemeinschaft under cluster of excellence ct.qmat (EXC 2147, Project-ID No. 390858490). This work was supported, in part, by the MIT–Poland Lockheed Martin Seed Fund under the MISTI Global Seed Funds program for the project Structural oddities.

{\it{Data availability}}$-$All code and data supporting this study are openly available \cite{Zenodo}.

\bibliography{references}

\newpage
\phantom{a}
\newpage
\setcounter{figure}{0}
\renewcommand{\thefigure}{S\arabic{figure}}
\renewcommand{\theHfigure}{S\arabic{figure}}
\setcounter{equation}{0}
\renewcommand{\theequation}{S\arabic{equation}}
\renewcommand{\theHequation}{S\arabic{equation}}
\setcounter{section}{0}

\onecolumngrid
\begin{center}
{\large \bf Supplementary Material:\\
Curved Odd Elasticity}\\
\vspace{0.3cm}
\end{center}
\setcounter{page}{1}

\section{Odd elasticity in curved background}
\label{sec:Derivation}

Elasticity on curved backgrounds can be introduced by employing two metrics, $\bar{g}_{ij}$ and $g_{ij}$. The first represents the reference metric, characterizing a strain- and stress-free configuration, while the second describes the medium after an arbitrary deformation in either shape or size. Accordingly, barred geometric quantities correspond to objects constructed from the reference metric, whereas unbarred quantities correspond to those constructed from the deformed metric.

We consider a two-dimensional elastic medium embedded in a curved spatial geometry described by the reference metric $\bar g_{ij}(x)$, where Latin indices run over $\{1,2\}$. The elastic deformation is parametrized by the displacement field $u_i(x,t)$. 

The strain tensor is defined from the linearized change of the induced spatial metric,
\begin{equation}
g_{ij}(x,t)=\bar g_{ij}(x)+2u_{ij}(x,t),
\end{equation}
valid to linear order in the deformation field.
The reference spatial metric takes the form
\begin{equation}
\bar g_{ij}
=
\begin{pmatrix}
\bar g_{11} & \bar g_{12} \\
\bar g_{21} & \bar g_{22}
\end{pmatrix}.
\end{equation}
We expand the elastic potential energy perturbatively around the undeformed configuration up to second
order in the strain tensor. Assuming a local elastic response,
\begin{equation}
W[\bar g_{ij}+2u_{ij}]
=
W[\bar g_{ij}]
+
\int d^2x \,
\sqrt{|\bar g|}
\left.
\frac{\delta W}{\delta g_{ij}}
\right|_{g=\bar g}
2u_{ij}
+
\frac12
\int d^2x d^2x'\,
\sqrt{|\bar g|}
\left.
\frac{\delta^2 W}
{\delta g_{ij}\delta g_{kl}}
\right|_{g=\bar g}
4u_{ij}u_{kl}
+\cdots .
\label{action}
\end{equation}
The zeroth-order contribution corresponds to the potential energy of the undeformed curved solid
and is independent of the displacement field. The first-order term vanishes when the reference configuration is an equilibrium configuration with no residual stress. More generally, this term encodes the pre-stress of the reference state and should be kept if the undeformed material is not stress free. The second-order term describes small elastic deformations around equilibrium and defines the rank-four elastic tensor in curved space. The time dependence enters through the displacement field $u^i(x,t)$. The induced spatial metric $g_{ij}[u]$ determines the elastic potential energy, while the kinetic energy is determined separately by $\partial_t u^i$. The quadratic action reads
\begin{equation}
S^{(2)}
=
\frac12
\int d^2xdt\,\sqrt{|\bar g|}
\left[
\rho\, \partial_t u_i \partial_t u^i
-
C^{ijkl}u_{ij}u_{kl}
\right],
\label{eq:2ndaction}
\end{equation}
where $\rho$ is the mass density, indices are raised using the reference metric
$\bar g^{ij}$, and the linear strain tensor is
\begin{equation}
u_{ij}
=
\frac12
\left(
\overline{\nabla}_i u_j
+
\overline{\nabla}_j u_i
\right).
\end{equation} $\overline{\nabla}_i$ denotes the covariant derivative compatible with the reference
metric $\bar g_{ij}$. Explicitly,
\begin{equation}
\overline{\nabla}_i u_j
=
\partial_i u_j
+
\overline{\Gamma}^k_{\,ij}u_k,
\end{equation}
where $\overline{\Gamma}^k_{\,ij}$ are the Christoffel symbols associated with the reference
metric $\bar g_{ij}$.

The isotropic elasticity tensor is given by
\begin{equation}
C^{ijkl}
=
(B-\mu)\bar g^{ij}\bar g^{kl}
+
\mu
\left(
\bar g^{ik}\bar g^{jl}
+
\bar g^{il}\bar g^{jk}
\right),
\label{elastictensor}
\end{equation}
where $B$ and $\mu$ are the bulk and shear moduli, respectively.
This tensor is symmetric under the exchange of the index pairs (Maxwell-Betti reciprocity). Its inverse is give by
\begin{equation}
(C^{abcd})^{-1}=\left(\frac{1}{4B}-\frac{1}{4\mu}\right)\bar{g}_{ab}\bar{g}_{cd}+\frac{1}{4\mu}(\bar{g}_{ac}\bar{g}_{bd}+\bar{g}_{ad}\bar{g}_{bc})=\frac{1}{Y}\left[\frac{(1+\nu)}{2}(\bar{g}_{ac}\bar{g}_{bd}+\bar{g}_{ad}\bar{g}_{bc})-\nu \bar{g}_{ab}\bar{g}_{cd}\right] .
\label{inverse}
\end{equation}
Furthermore
\begin{equation}
C^{abcd}C_{cdef}=\frac{1}{2}(\bar{g}^a_e\bar{g}^b_f+\bar{g}^a_f\bar{g}^b_e) ,
\end{equation}
where $Y=2 B (1-\nu)=2\mu (1+\nu)$ is the Young's modulus and $\nu=(B-\mu)/(B+\mu)$ is the two-dimensional Poisson ratio ranging from $-1$ for auxetic materials to 1 for fully incompressible solids. 
Variation of the action in Eq.~\eqref{eq:2ndaction} with respect to the strain tensor and the time derivative of the displacement field yields the stress tensor and the momentum density, respectively
\begin{equation}
\begin{aligned}
\sigma^{ab} &=\frac{1}{\sqrt{|\Bar{g}|}}\frac{\delta S}{\delta u_{ab}}=C^{abcd}u_{cd}, \\
\pi^a &=\frac{1}{\sqrt{|\Bar{g}|}}\frac{\delta S}{\delta (\partial_t u_a)}=\rho \partial_t u^a .
\end{aligned}
\end{equation}
The equation of motion follows from variation of the action
\begin{equation}
\delta S=\int \sqrt{|\bar g|}
\left[
\rho \partial_t^2 u^a-
\bar\nabla_b \sigma^{ab}
\right]
\delta u_a \,
d^2xdt,
\end{equation}

The next step is the extension of the rank-4 tensor with odd elastic terms related to the modulus $K_o$. In principle, the inclusion of the odd modulus $A$ is also possible for a solid with internal torque density, but for simplicity we choose not. This term in a curved background is given by:
\begin{equation}
C_{odd}^{abcd}=\frac{K_o}{2}\left(\Bar{g}^{ac}\bar{\epsilon}^{bd}+\Bar{g}^{ad}\bar{\epsilon}^{bc}+\Bar{g}^{bc}\bar{\epsilon}^{ad}+\Bar{g}^{bd}\bar{\epsilon}^{ac}\right)
\end{equation}
where in our case $\bar{\epsilon}^{ab}=\epsilon^{ab}/\sqrt{|\Bar{g}|}$ \footnote{If we include the odd contribution before the covariant derivative bypasses the rank-4 elasticity tensor during the extraction of the equation of motion, it would still have no difference in the final result as $\overline{\nabla}_{\alpha}\sqrt{|\Bar{g}|}=0$.}. The final result is the odd Navier's equation for the displacement field
\begin{equation}
\rho \frac{\partial^2 u_i}{\partial t^2}
+\gamma \frac{\partial u_i}{\partial t}
-\overline{\nabla}_i \left[
(B-\mu)\left(\overline{\nabla}_ju_j\right)
+K_o\left(\bar{\epsilon}_{jk}\overline{\nabla}_ju_k\right)
\right]
-\overline{\nabla}_j^2 \left[
2\mu u_i+K_o(\bar{\epsilon}_{ik}u_k)
\right]
-K_o\bar{\epsilon}_{ik}\overline{\nabla}_k\left(\overline{\nabla}_ju_j\right)=0  
\end{equation}
where we introduce a damping effect with positive parameter $\gamma$.
The structure of the above second-order, coupled linear equations makes finding closed-form solutions particularly challenging. To address this, we introduce two scalar functions (Papkovich–Neuber potentials):
\begin{equation}
u_i=\overline{\nabla}_i\chi+\bar{\epsilon}_{ij}\overline{\nabla}^j\psi.
\label{Papkovich-Neuber}
\end{equation}
Substituting this representation into the governing equations yields the following form:
\begin{equation*}
\overline{\nabla}_i
\left[\rho\frac{\partial^2 \chi}{\partial t^2}+\gamma\frac{\partial \chi}{\partial t}-(B+\mu)\overline{\nabla}^2\chi+K_o\overline{\nabla}^2\psi\right]+\bar{\epsilon}_{ik}\overline{\nabla}^k
\left[\rho\frac{\partial^2 \psi}{\partial t^2}+\gamma\frac{\partial \psi}{\partial t}-\mu\overline{\nabla}^2\psi-K_o\overline{\nabla}^2\chi\right]=0 .
\end{equation*}
This expression can be satisfied only if \footnote{We clarify that every solution of Navier's equation admits a representation like the one provided by Eq.~\eqref{Papkovich-Neuber}.}:
\begin{equation}
\begin{aligned}
    \rho\frac{\partial^2 \chi}{\partial t^2}&+\gamma\frac{\partial \chi}{\partial t}-(B+\mu)\overline{\nabla}^2\chi+K_o\overline{\nabla}^2\psi=0,\\
    \rho\frac{\partial^2 \psi}{\partial t^2}&+\gamma\frac{\partial \psi}{\partial t}-\mu\overline{\nabla}^2\psi-K_o\overline{\nabla}^2\chi=0.
\label{EOM1}
\end{aligned}
\end{equation}
These coupled linear differential equations written in general curvilinear coordinates also provide the usual solutions for the modes for the case of 2D flat surface (infinitely long plate). In the case where $K_o=0$, we also recover back the usual solutions for the passive elastic solid. In the next sections we will be applying all our results to the cases of a spherical and a toroidal surface.

\section{Application on the 2-sphere}
\label{sec:sphere}
For the case of a 2-sphere of radius $r$, polar angle $\theta$ and azimuthal angle $\phi$, 
\begin{equation}
    \begin{aligned}
        x = &r \sin\theta \cos\phi, \\
        y = &r \sin\theta \sin\phi, \\
        z = &r \cos\theta.
    \end{aligned}
\end{equation}
where we want to apply our findings, we have
\begin{equation}
\begin{aligned}
\overline{g}_{ab} &= 
\begin{pmatrix}
r^2 & 0\\
0 &  r^2 \sin^2\theta\\
\end{pmatrix}, \\
\overline{g}^{ab}&= 
 \begin{pmatrix}
\frac{1}{r^2} & 0\\
0 & \frac{1}{r^2 \sin^2\theta}\\
\end{pmatrix}  .
\end{aligned}
\end{equation}
At the same time for the Ricci scalar and the square root of the metric determinant
\begin{equation}
\mathcal{R}=\frac{2}{r^2},  \hspace{1cm} \sqrt{|\Bar{g}|}=r^2\sin\theta .
\end{equation}

By making use of the Laplace--Beltrami operator on a two-dimensional sphere of fixed radius $r$,
\begin{equation}
\overline{\nabla}^2=
\frac{1}{\sqrt{|\Bar{g}|}}\partial_a(\sqrt{|\Bar{g}|}\bar{g}^{ab}\partial_b)
=\frac{1}{r^2}\left[
\frac{\partial^2}{\partial\theta^2}
+\cot\theta\,\frac{\partial}{\partial\theta}
+\frac{1}{\sin^2\theta}\frac{\partial^2}{\partial\phi^2}
\right],
\end{equation}
and expanding the scalar potentials in spherical time harmonics,
\begin{equation}
\begin{aligned}
\chi(\theta,\phi,t)
&=\sum_{l=0}^{\infty}\sum_{m=-l}^{l}
C_{lm}\,Y_l^{m}(\theta,\phi)\,e^{-i\omega t},\\[4pt]
\psi(\theta,\phi,t)
&=\sum_{l=0}^{\infty}\sum_{m=-l}^{l}
D_{lm}\,Y_l^{m}(\theta,\phi)\,e^{-i\omega t},
\end{aligned}
\label{sphericalharmonics}
\end{equation}
where $C_{lm}$ and $D_{lm}$ depend on the mode numbers $(l,m)$. We use the eigenvalue relation
\begin{equation}
r^{2}\,\overline{\nabla}^{2} Y_l^{m}(\theta,\phi)
= -\,l(l+1)\,Y_l^{m}(\theta,\phi),
\end{equation}
together with the completeness relation
\begin{equation}
\int_{0}^{2\pi}\int_{0}^{\pi}
Y_l^{m}(\theta,\phi)\,
\big(Y_{l'}^{m'}(\theta,\phi)\big)^{*}
\sin\theta\, d\theta\, d\phi
= \delta_{l,l'}\,\delta_{m,m'}.
\end{equation}
Projecting the governing equations onto this orthonormal basis yields, for each mode $(l,m)$, the $2\times 2$ matrix system
\begin{equation}
\begin{pmatrix}
\rho\,\omega^{2} + i\gamma\omega - \dfrac{(B+\mu)}{r^{2}}\,l(l+1)
&
\dfrac{K_o}{r^{2}}\,l(l+1)
\\[10pt]
-\dfrac{K_o}{r^{2}}\,l(l+1)
&
\rho\,\omega^{2} + i\gamma\omega - \dfrac{\mu}{r^{2}}\,l(l+1)
\end{pmatrix}.
\end{equation}

Setting the determinant of the above matrix equal to zero yields a quartic
(characteristic polynomial of rank~4) of the form
\begin{equation}
\omega^{4} + c_3\,\omega^{3} + c_2\,\omega^{2} + c_1\,\omega + c_0 = 0,
\end{equation}
where
\begin{equation}
\begin{aligned}
c_3 &= \,2i\,\frac{\gamma}{\rho}, \\
c_2 &= -\,\frac{\rho(B+2\mu)\,l(l+1) + \gamma^{2} r^{2}}{\rho^{2} r^{2}}, \\
c_1 &= -i\,\frac{\gamma (B+2\mu)\,l(l+1)}{\rho^{2} r^{2}}, \\
c_0 &= \frac{\mu(B+\mu) + K_o^{2}}{\rho^{2} r^{4}}\,l^{2}(l+1)^{2}.
\end{aligned}
\end{equation}
Solving this quartic for the frequency yields the four eigenvalues
\begin{equation}
\omega = -i\,\frac{\gamma}{2\rho}\;
\pm\;\sqrt{
\frac{2\rho\left(B+2\mu \pm \sqrt{\,B^{2} - 4K_o^{2}\,}\right)\,l(l+1)- \gamma^{2}r^{2}}
{4\rho^2 r^2}}.
\end{equation}
In the limit $\gamma=0$, the solutions come out as 
\begin{equation}
\lambda_{\pm} = \rho \omega^2_{\text{inertia}}=\frac{B+2\mu\pm\sqrt{B^2-4K_o^2}}{2}\frac{l(l+1)}{r^2}
\label{eq:sphere_inertial}
\end{equation}
The spectrum is inversely proportional to $r^2$. The flat limit for the 2D infinitely long odd plate can in principle be recovered if the fraction of $l^2/r^2$ for large $l$ is instead replaced for the $q^2$ of the wavenumber for the linear spectrum of the continuum limit.

In the overdamped regime on the other hand ($\rho \ll \gamma$), the solution is
\begin{equation}
\lambda_{\pm} = i\gamma \omega_{\text{overdamped}}=\frac{B+2\mu\pm\sqrt{B^2-4K_o^2}}{2} \frac{l(l+1)}{r^2}.
\label{eq:sphere_overdamped}
\end{equation}
The spectrum now is inversely proportional to $r^2$ and quadratically dependent on the quadratic in the angular mode index $l$. It is clear from Eq.~\eqref{eq:sphere_overdamped} that just like the infinitely long plate theory, wave solutions can still be present in the overdamped case where $4K_o^2>B^2$ as now $\omega_{\text{overdamped}}$ acquires a real part.

The corresponding eigenvector is
\begin{equation}
\begin{pmatrix}
C_{lm} \\[4pt]
D_{lm}
\end{pmatrix}
=
\begin{pmatrix}
B + \sqrt{B^{2} - 4K_o^{2}} \\
2K_o
\end{pmatrix}
,
\begin{pmatrix}
2K_o \\
B + \sqrt{B^{2} - 4K_o^{2}}
\end{pmatrix}.
\end{equation}

\section{Application on a toroidal surface}
\label{sec:torus}
The torus formed when rotating around the axis $z$, a ring of radius $r$, whose center lies in the $xy$ plane, a distance $R$ from point $(0,0,0)$, can be parametrized as
\begin{equation}
\begin{aligned}
x &= (R+r \cos \theta) \cos \phi ,\\
y &= (R+r \cos \theta) \sin \phi ,\\
z &= r \sin \theta,
\end{aligned}
\end{equation}
where $R>0$, $r>0$, $-\pi\leq \phi, \theta \leq \pi$,
or
\begin{equation}
\begin{aligned}
x &= a \frac{\sinh{\tau}}{\cosh{\tau}-\cos{\sigma}} \cos{\phi} ,\\
y &= a \frac{\sinh{\tau}}{\cosh{\tau}-\cos{\sigma}} \sin{\phi}, \\
z &= a \frac{\sin{\sigma}}{\cosh{\tau}-\cos{\sigma}},
\end{aligned}
\end{equation}
where the coordinate ranges are $a> 0$, $\tau > 0$, $-\pi \leq \phi,\sigma \leq \pi$. 
The geometric parameters are
\begin{equation}
\begin{aligned}
    R&=a\coth{\tau}, \quad r=a \csch{\tau}, \\
    a&=\sqrt{R^2-r^2}, \quad \tau= \log\left(\frac{R}{r}+\sqrt{\frac{R^2}{r^2}-1}\right).
\end{aligned}
\end{equation}
The area of a torus is $A=4\pi^2 R r=4\pi^2 a^2 \coth{\tau} \csch{\tau}$.

We choose to work here with the second option for the modal solutions. The metric tensor and its inverse for a toroid (also called a 2-Torus), is given by
\begin{equation}
\begin{aligned}
\overline{g}_{ij}&=
\begin{pmatrix}
\frac{a^2 \sinh^2{\tau}}{\left(\cosh{\tau}-\cos{\sigma}\right)^2} & 0 \\
0 & \frac{a^2}{\left(\cosh{\tau}-\cos{\sigma}\right)^2}
\end{pmatrix}, \\
\overline{g}^{ij}&=
\begin{pmatrix}
\frac{\left(\cosh{\tau}-\cos{\sigma}\right)^2}{a^2 \sinh^2{\tau}} & 0 \\
0 & \frac{\left(\cosh{\tau}-\cos{\sigma}\right)^2}{a^2 }
\end{pmatrix},
\end{aligned}
\end{equation}
and appropriately, we can define the Laplace-Beltrami operator
\begin{equation}
    \overline{\nabla}^2 = 
    \frac{\left(\cosh{\tau}-\cos{\sigma}\right)^2}{a^2}
    \left(
    \frac{1}{\sinh^2{\tau}}\frac{\partial^2 }{\partial \phi^2}
    +\frac{\partial^2 }{\partial \sigma^2} 
    \right).
\end{equation}

To make things simpler, we will choose to work on a toroid where the radius from the center is much bigger than the internal radius of the ring, i.e. $R\gg r$. In that approximation, the $\cos\sigma$, can be dropped and we can use the simple ansatz for the general solution of the scalar potentials 
\begin{equation}
\begin{aligned}
 \chi(\phi,\sigma) &\sim \Tilde{\chi} e^{im\phi}e^{in\sigma} ,\\
 \psi(\phi,\sigma) &\sim \Tilde{\psi} e^{im\phi}e^{in\sigma} .
\end{aligned}
\end{equation}
The eigenvalue problem takes the form ($\lambda = \rho \omega^2_{\text{inertia}}$ or $\lambda = i\gamma \omega_{\text{overdamped}}$)
\begin{equation}
-\lambda
    \begin{pmatrix}
    \Tilde{\chi} \\ \Tilde{\psi}
    \end{pmatrix}
=-\frac{\cosh^2{\tau}}{a^2}
    \left(\frac{m^2}{\sinh^2{\tau}}+n^2\right)
    \begin{pmatrix}
        B+\mu & -K_o \\
        K_o & \mu
    \end{pmatrix}
    \begin{pmatrix}
    \Tilde{\chi} \\ \Tilde{\psi}
    \end{pmatrix},
\end{equation}
with spectrum
\begin{equation}
    \lambda_{\pm} = \frac{B+2\mu\pm \sqrt{B^2-4K_o^2}}{2} \frac{\cosh^2{\tau}}{a^2}
    \left(\frac{m^2}{\sinh^2{\tau}}+n^2\right).
\label{eq:torus_lambda}
\end{equation}
In the limit $R/r \to \infty$, the equation reduces to that of a cylindrical surface:
\begin{equation}
    \lambda_{\pm} = \frac{B+2\mu\pm \sqrt{B^2-4K_o^2}}{2} 
    \frac{n^2}{r^2}.
\end{equation}

The eigenvector is 
\begin{equation}
    \begin{pmatrix}
    \Tilde{\chi} \\ \Tilde{\psi}
    \end{pmatrix}
    =
    \begin{pmatrix}
    B + \sqrt{B^2-4K_o^2} \\ 2K_o
    \end{pmatrix},
    \begin{pmatrix}
     2K_o \\ B + \sqrt{B^2-4K_o^2}
    \end{pmatrix}
\end{equation}
The displacement fields Eq.~\eqref{Papkovich-Neuber} read
\begin{equation}
\mathbf{u}= \left(u_\phi, u_\sigma \right)=
\frac{\cosh{\tau}-\cos{\sigma}}{a}
\left( 
\frac{1}{\sinh\tau} \frac{\partial \chi}{\partial \phi}+ 
\frac{\partial \psi}{\partial \sigma}
, \frac{\partial \chi}{\partial \sigma} 
-\frac{1}{\sinh\tau}  \frac{\partial \psi}{\partial \phi}
\right),
\end{equation}
where the two components correspond to the $\mathbf{e}_\phi$ and $\mathbf{e}_\sigma$ directions on the torus. The strain fields are
\begin{equation}
    \overline{\nabla} \mathbf{u} =
    \begin{pmatrix}
        \frac{\cosh{\tau}-\cos{\sigma}}{a \sinh{\tau}} \frac{\partial u_\phi}{\partial \phi} - \frac{\sin\sigma}{a} u_\sigma &  
        \frac{\cosh{\tau}-\cos{\sigma}}{a \sinh{\tau}} \frac{\partial u_\sigma}{\partial \phi} + \frac{\sin\sigma}{a} u_\phi \\
        \frac{\cosh{\tau}-\cos{\sigma}}{a} \frac{\partial u_\phi}{\partial \sigma} & 
        \frac{\cosh{\tau}-\cos{\sigma}}{a} \frac{\partial u_\sigma}{\partial \sigma}
    \end{pmatrix}
\end{equation}

\section{Flat lattices}\label{sec:Microscopics}
\begin{figure}[t!]
    \centering
    \includegraphics[width=0.7 \linewidth]{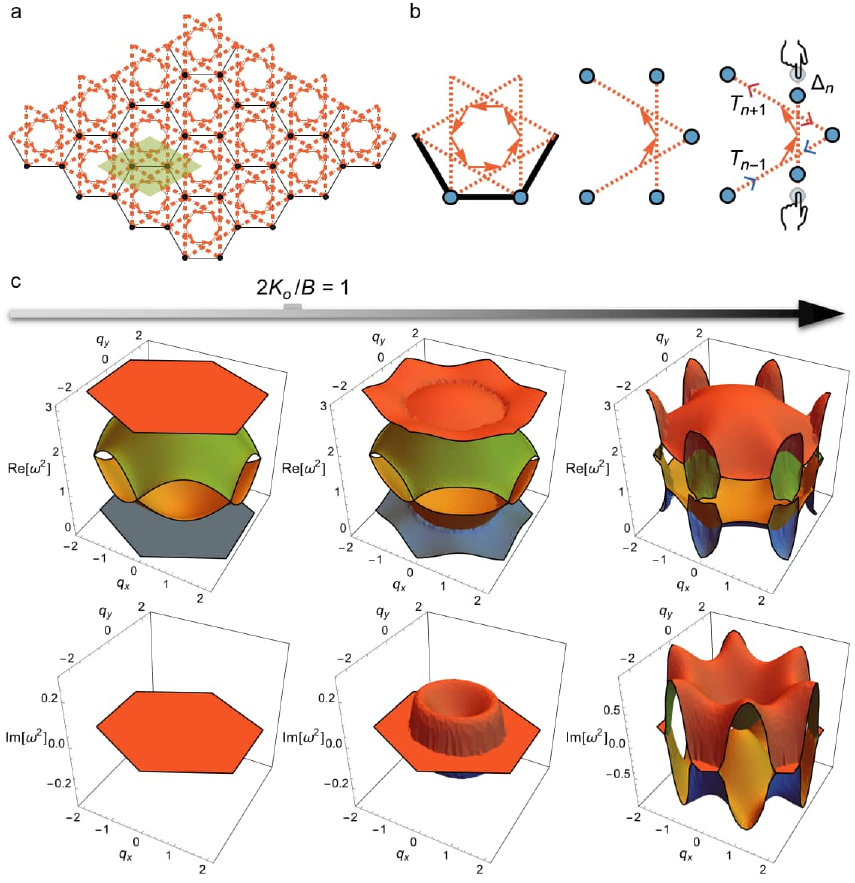 }
    \caption{Band structure of a flat honeycomb lattice. 
    (a) A honeycomb lattice.
    (b) The unit cell with nonreciprocal bonds obeying the tension-length relationship $T_n=k_o(\Delta_{n+1}- \Delta_{n-1})$. Here $T_n$ is the tension in bond $n$, and $\Delta_{n+1}$, $ \Delta_{n-1}$ are the length changes in the neighboring bonds $n+1$ and $n-1$ within a hexagonal plaquette.
    An example deformation, in which bond $n$ is compressed with length change $\Delta_n < 0$, causing the neighboring bonds to experience tension $T_{n+1} = -k_o \Delta_n$ and compression $T_{n-1} = k_o \Delta_n$. 
    (c) The spectrum plotted over the Brillouin zone for $k_o/k=0$, $0.16$, $0.4$.
    }
    \label{fig:bands}
\end{figure}

A lattice is described by a set of nodes connected by bonds.
Here we consider a flat honeycomb lattice [Figs.~\ref{fig:bands}(a, b)] with passive nearest- and next-nearest-neighbor interactions characterized by Hookean spring stiffness $k$ and $k'$, together with nonreciprocal interactions characterized by
\begin{equation}
    T_n=k_o (\Delta_{n+1} - \Delta_{n-1}), 
\end{equation}
where $\Delta_n$, $T_n$ denote bond elongation and tension, $k_o$ represent reciprocal and nonreciprocal stiffness. 
The bonds within each hexagonal plaquette are numbered counterclockwise, and bonds shared by two neighboring plaquettes contribute to both plaquettes. 
Active tensions are generated through the nonreciprocal couplings. For example, when bond $n$ is compressed with length change $\Delta_n<0$, the neighboring bonds experience tension $T_{n+1} = -k_o \Delta_n$ and compression $T_{n-1} = k_o \Delta_n$ [Fig.~\ref{fig:bands}(b)].

The linear dynamics are governed by the equation of motion
\begin{equation}
    \mathbf{M} \Ddot{\mathbf{u}} =\mathbf{F} = -\mathbf{D} \mathbf{u}
    = -\mathbf{C}^T \mathbf{K} \mathbf{C} \mathbf{u}, 
    \label{eq: EOM of lattices}
\end{equation}
where $\mathbf{M}$ is the diagonal matrix of nodal mass, and $\mathbf{D}$ is the dynamical matrix.
The geometric information is encoded by the  compatibility matrix $\mathbf{C}$, and the elastic interactions are assembled into the stiffness matrix $\mathbf{K}$. It is a diagonal matrix $diag(k_1,k_2,\dots,k_{N_b},k'_1,k'_2,\dots,k'_{NN_b})$ in reciprocal systems, while nonreciprocal interactions introduce antisymmetric terms $k_o$.

For periodic lattices, we can take the Fourier transform of Eq.~\eqref{eq: EOM of lattices}
\begin{equation}
    -\mathbf{M} \omega^2 \Tilde{\mathbf{u}}=-C^{\dagger}(\mathbf{q}) \mathbf{K} C(\mathbf{q})  \Tilde{\mathbf{u}},
\end{equation}
where $\mathbf{q}$ is the wave-vector. The band structure can be solved from the equation by calculating the eigenvalues for each wave-vector at the Brillouin zone.  We find the real bands become complex at a critical nonreciprocal stiffness [Fig.~\ref{fig:bands}(c)], a typical feature of exceptional transition. Interestingly, the instability appears first at the long-wavelength limit and gradually immerse the whole Brillouin zone.

Following a standard coarse graining procedure \cite{lubensky2015Phonons,binysh2025More}, we extract the isotropic elastic modulus tensor for the nonreciprocal honeycomb lattice. 
We choose the four basis for second-order tensor
\begin{equation}
    \tau^0 =
    \begin{pmatrix}
        1 & 0 \\ 0 & 1
    \end{pmatrix}
    ,\quad
    \tau^1 =
    \begin{pmatrix}
        0 & -1 \\ 1 & 0
    \end{pmatrix}
    ,\quad
    \tau^2 =
    \begin{pmatrix}
        1 & 0 \\ 0 & -1
    \end{pmatrix}
    ,\quad
    \tau^3 =
    \begin{pmatrix}
        0 & 1 \\ 1 & 0
    \end{pmatrix},
\end{equation}
and the constitutive equation is expressed in the matrix form
\begin{equation}
    \begin{pmatrix}
        \sigma^0 \\ \sigma^1 \\ \sigma^2 \\ \sigma^3 
    \end{pmatrix}
    =2
    \begin{pmatrix}
    B & 0 & 0 & 0 \\
    0 & 0 & 0 & 0 \\
    0 & 0 & \mu & K_o \\
    0 & 0 & -K_o & \mu
    \end{pmatrix}
    \begin{pmatrix}
      u^0 \\ u^1 \\ u^2 \\ u^3  
    \end{pmatrix}
\end{equation}
where bulk modulus, shear modulus, and odd modulus are
\begin{equation}
    B=\frac{k+6k'}{2\sqrt{3}},\quad \mu=\frac{\sqrt{3}k'}{2},\quad K_o=\frac{3k_o}{2}.
\end{equation}

\begin{figure}[b!]
    \centering
    \includegraphics[width=0.7 \linewidth]{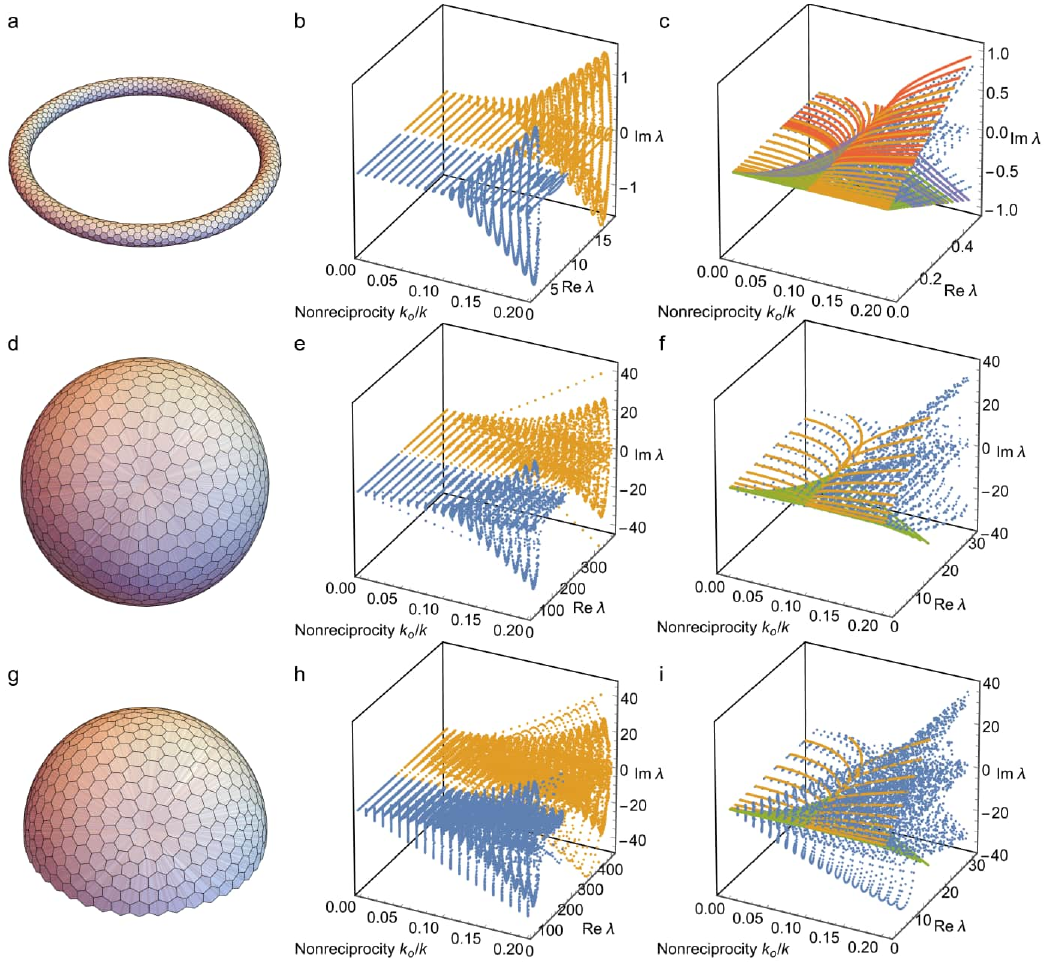}
    \caption{Vibrational spectra of curved systems. 
    (a, d, g) A torus, full spherical surface, and a half-spherical cap.
    (b, e, h) The $2N$ in-plane acoustic (blue) and optical (yellow) branches.
    (c, f, i) The low-frequency spectra and theoretical predictions of curved odd elasticity. 
    (c) The yellow and orange branches ($m=1,2,\dots$ in Eq.~\eqref{eq:torus_lambda}) denote $\lambda_+$ with $n=0$ and $n=1$. The green and purple branches denote $\lambda_-$ with $n=0$ and $n=1$.
    (f, i) The yellow and green branches ($l=1,2,\dots$ in Eq.~\eqref{eq:sphere_inertial}) denote $\lambda_+$ and $\lambda_-$, respectively.
    }
    \label{fig:all_data}
\end{figure}

\section{Curved systems}
\label{sec: curved}
For our curved systems, we introduce another strong geometric constraints $k_c$ that confine nodal displacements to the tangent plane. Specifically, we introduce springs with stiffness $k_c$ to connect each node $(x,y,z)$ on surfaces to the axial circle $(R\cos\phi,R\sin\phi,0)$ of a torus or the origin $(0,0,0)$ of a spherical surface. In the following analysis, we set the stiffness $k_c/k=10^4$ to impose strong constraints, $k'/k=10^{-4}$ to introduce nonzero shear modulus, and $k_o/k \sim 1$ to capture the odd effects. 

For a curved system consisting of $N$ nodes, there are $3N$ degrees of freedom. The out-of-plane responses exhibit $N$ high frequencies $\omega^2 \sim k_c/m_a$, and the in-plane motions are composed with the optical and acoustic branches, as the band structure of the diatomic honeycomb lattice shown in Figs.~\ref{fig:bands}(b, c). Furthermore, the acoustic branch consists of longitudinal $\omega^2 \sim k/m_a$ and transverse $\omega^2 \sim k'/m_a$ sections in passive systems, which are involved with the bulk and shear moduli in the continuum limit. The existence of nonreciprocal interactions mix the longitudinal and transverse components and therefore lead to chiral phonons. The vibrational spectra of curved system are shown in Fig.~\ref{fig:all_data}. To compare with the analytical predictions ($\lambda = \rho \omega^2_{\text{inertia}}$), we calculate the density of each curved lattice $\rho = Nm_a/A$, where $A$ is the area of a torus $4\pi^2 Rr$, a spherical surface $4\pi r^2$, or a half-spherical cap $2\pi r^2$.

To characterize energy injection in curved systems, we evaluate the power associated with each eigenmode. For a reciprocal system, the power is always zero.
\begin{equation}
P=\operatorname{Re}\braket{\dot{\mathbf{u}}|\mathbf{F}}
=\operatorname{Re}(-i \omega \mathbf{u}^\dagger \mathbf{C}^T \mathbf{K} \mathbf{C} \mathbf{u})
\propto \operatorname{Im}(\mathbf{u}^\dagger \mathbf{C}^T \mathbf{K_o} \mathbf{C} \mathbf{u})
\end{equation}
 While in nonreciprocal systems, the stiffness matrix contains antisymmetric components. These antisymmetric contributions generate nonzero local power on each plaquette, so that the total power can be written as $P=\sum_{\text{plaquette}} P_i$. We therefore color each hexagonal units to visualize the distribution and identify power reversals. To globally measure the degree of power reversal in each eigenmode, we sum up the positive $P^+=\sum (P_i>0)$ and negative $P^-=\sum (P_i<0)$ power contributions and introduce the power ratio $p=|P^{-}|/|P^{+}|$.

\subsection{Defect modes}
\begin{figure}[ht!]
\includegraphics[width=0.70\linewidth]{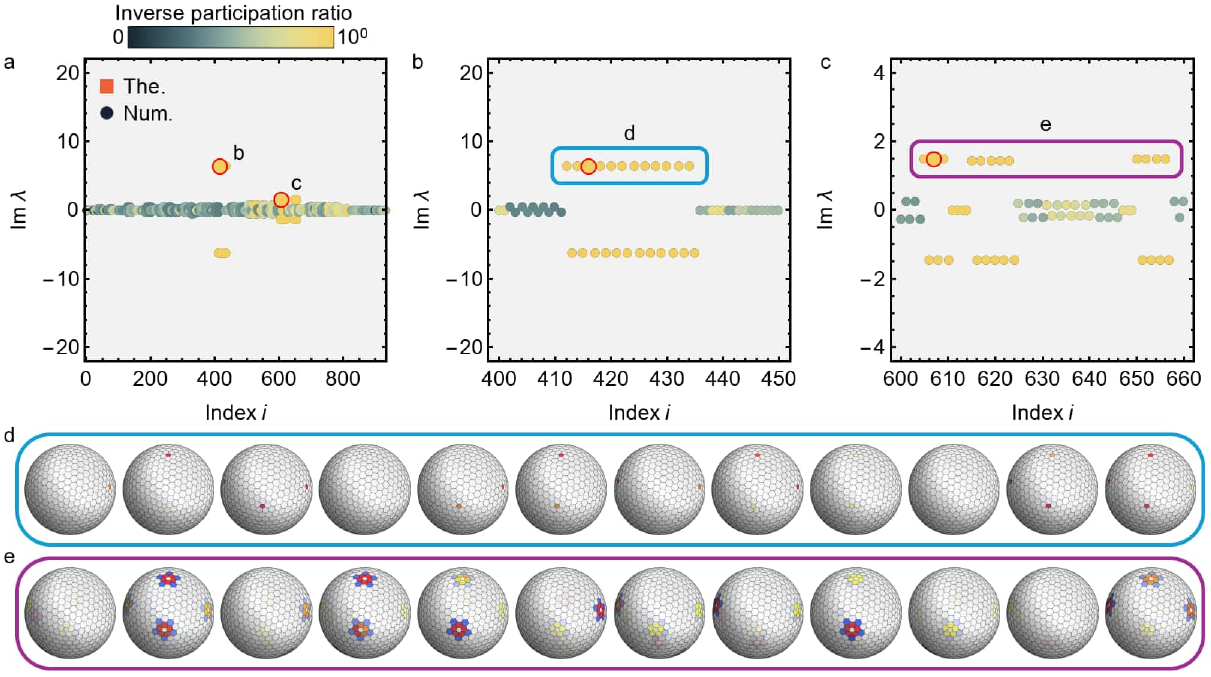}
    \caption{
    {\bf Spherical defect modes.}
   (a) Eigenvalues ordered by their real parts.
   (b, c) Zoom-in views of the unstable defect modes.
   (d, e) Degenerate defect modes.
   }
   \label{fig: full degenerate defect}
\end{figure}

\begin{figure}[h!]
\includegraphics[width=0.70\linewidth]{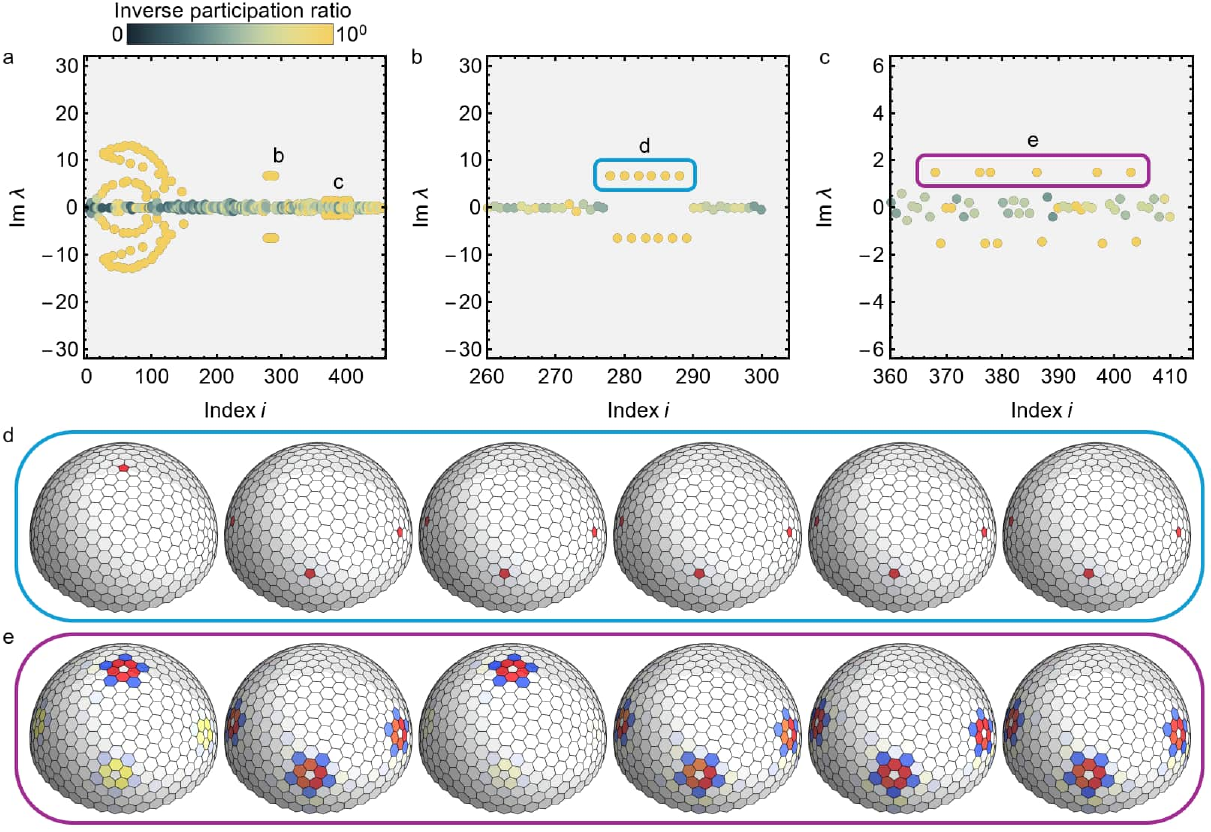}
    \caption{
    {\bf Spherical-cap defect modes.}
   (a) Eigenvalues ordered by their real parts.
   (b, c) Zoom-in views of the unstable defect modes.
   (d, e) Degenerate defect modes.
   }
   \label{fig: half degenerate defect}
\end{figure}

The number of localized defect modes equals the number of topological defects. For the full spherical lattices in Fig.~\ref{fig: full degenerate defect}, there are 12 pentagons and correspondingly 12 defect modes. For the spherical cap in Fig.~\ref{fig: half degenerate defect}, there are 6 pentagons and 6 defect modes. These modes remain largely localized at the defects, and their weak hybridization with bulk modes preservers their degeneracy.


\subsection{Edge modes}
\begin{figure}[ht!]
\includegraphics[width=0.75\linewidth]{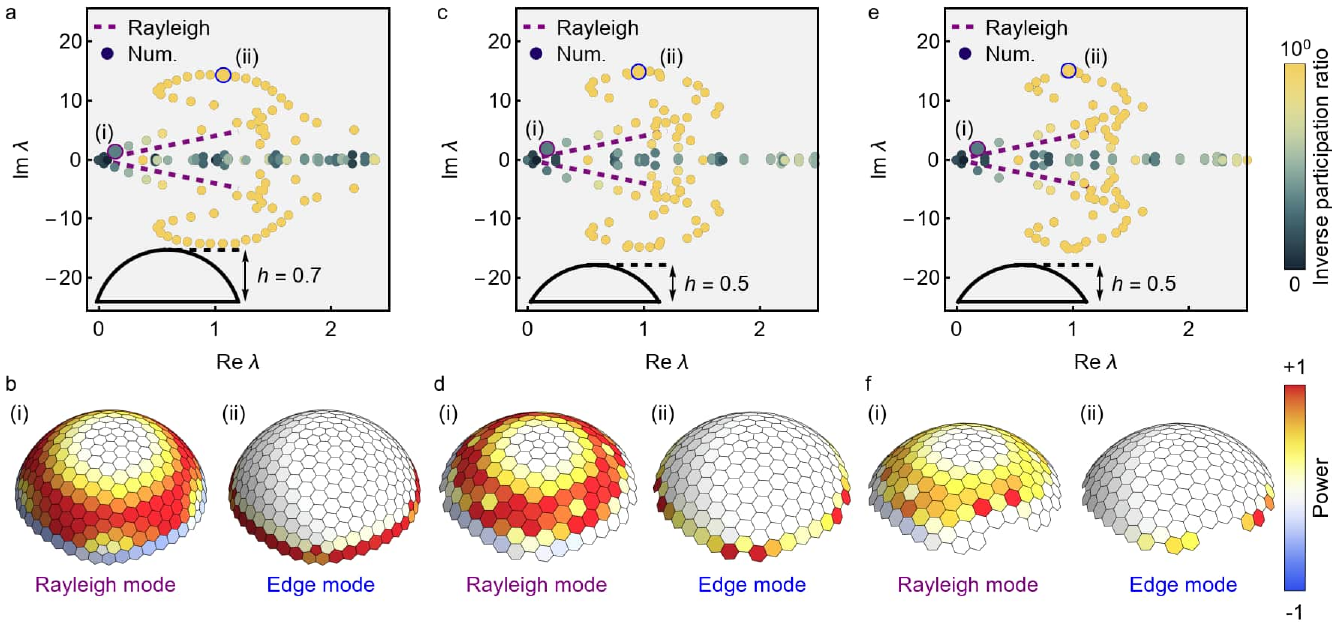}
    \caption{
    {\bf Robust edge modes.}
   (a, b) A spherical cap with height $h=0.7$ exhibits (i) Rayleigh modes with finite penetration depth and (ii) lattice-scale edge modes.
   (c, d) A spherical cap with height $h=0.5$ still supports (i) Rayleigh modes with finite penetration depth and (ii) lattice-scale edge modes. 
   (e, f) A spherical cap with height $h=0.5$ after manually removing three edge units shows (i) Rayleigh modes and (ii) lattice-scale defective edge modes.
    }
   \label{fig: robust edge modes}
\end{figure}

To investigate the effects of boundary geometry on the edge modes, we consider three spherical caps ($r=1$): a cap with height $h=0.7$ [Fig.~\ref{fig: robust edge modes}(a, b)], a cap with height $h=0.5$ [Fig.~\ref{fig: robust edge modes}(c, d)], and a cap with manually removed edge units [Fig.~\ref{fig: robust edge modes}(e, f)]. In all cases, we observe (i) Rayleigh modes with finite penetration depth and (ii) lattice-scale edge modes. 
These results demonstrate that the boundary geometry, including both the number of units and the boundary shape, has only a minor influence on the eigenspectra of the Rayleigh and edge modes, although it can affect the degree of mode localization.

\subsection{Potential signatures in embryonic crystals}

\begin{figure}[th!]
\includegraphics[width=0.5\linewidth]{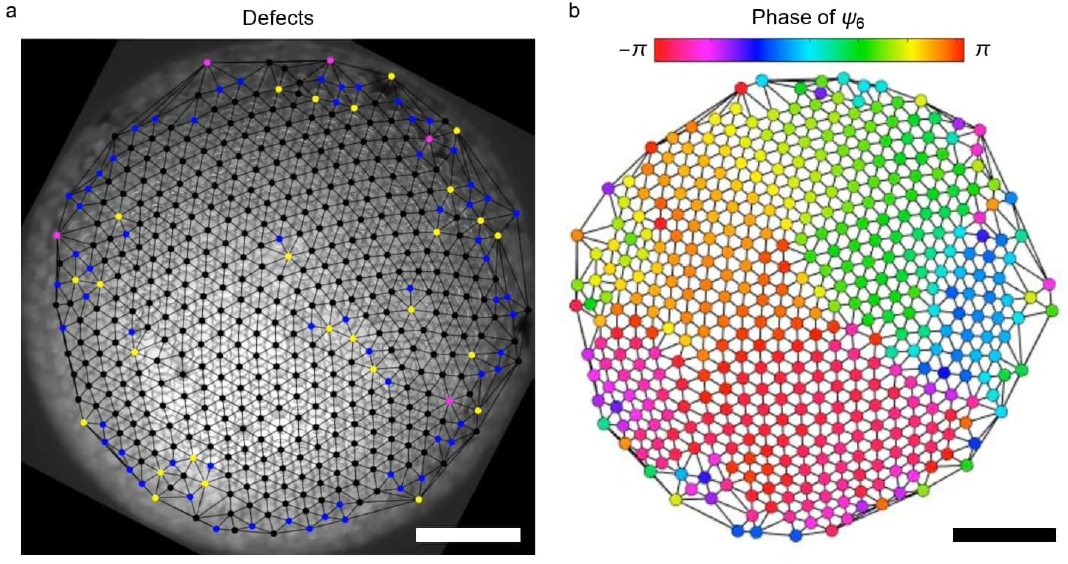}
   \caption{ 
   (a) Spinning embryos in the crystal form a hexagonal lattice, containing fivefold (blue) and seven (yellow) fold defects.
   (b) Phase of $\psi_6$. Scale bar, 1 mm.
    }
   \label{fig: phase}
\end{figure}

Our predictions are directly testable in living and engineered curved active solids. In living chiral crystals assembled on curved interfaces [Fig.~\ref{fig: phase}], geometry and topology enforce defects and boundaries that qualitatively reorganize the dynamics. In particular, the disclinations imposed by curvature in Fig.~\ref{fig: phase} are expected to act as localized sources of oscillations, activating at small non-reciprocity and entraining the surrounding collective.
 
These signatures are accessible via particle image velocimetry and single-particle tracking methods developed for living chiral crystals \cite{tan2022Odd, chao2026Selective}. A direct test would be the comparison of measured displacement fields $|u(x)|$ with the predicted eigenmodes, as in Fig.~2(g, h); because the spatial structure is primarily set by geometry, this comparison depends only weakly on material parameters.

\section{Rayleigh waves}
\label{sec: Rayleigh}
Here we derive the dispersion relation for Rayleigh waves in odd elastic systems.
We start from the linear odd elastodynamic equation Eq.~\eqref{EOM1} in a flat system and, for convenience, consider the overdamped case ($\rho=0$).
For a bulk medium occupying the half-plane ($y<0$), we assume scalar potentials of the form
\begin{equation}
\begin{aligned}
 \chi  &= \Tilde{\chi} e^{iq_x x}e^{k_y y}e^{-i\omega t} ,\\
 \psi  &= \Tilde{\psi} e^{iq_x x}e^{k_y y}e^{-i\omega t}  .
\end{aligned}
\end{equation}
where $k_y$ is the decay factor.
Substituting these solutions into the equations of motion Eq.~\eqref{EOM1}, we obtain the following nontrivial solutions 
\begin{equation}
\begin{aligned}
    k_l^2 &= q_x^2 - \frac{2 i \gamma \omega }{B+2\mu+\sqrt{B^2-4K_o^2}} \\ 
    \begin{pmatrix}
    \Tilde{\chi}_l \\ \Tilde{\psi}_l
    \end{pmatrix}
    &=
    \begin{pmatrix}
    B + \sqrt{B^2-4K_o^2} \\ 2K_o
    \end{pmatrix}
\end{aligned}
\label{eq: kl}
\end{equation}
and
\begin{equation}
\begin{aligned}
    k_t^2 &= q_x^2 - \frac{2 i \gamma \omega}{B+2\mu-\sqrt{B^2-4K_o^2}} \\ 
    \begin{pmatrix}
    \Tilde{\chi}_t \\ \Tilde{\psi}_t
    \end{pmatrix}
    &=
    \begin{pmatrix}
    2K_o \\ B + \sqrt{B^2-4K_o^2}
    \end{pmatrix} .
\end{aligned}
\label{eq: kt}
\end{equation}
We now combine these two solutions to construct Rayleigh waves.
The displacement field is given by Eq.~\eqref{Papkovich-Neuber}, which yields
\begin{equation}
    \mathbf{u} = \left\{ m_1 e^{k_l y} 
    \left[ \tilde{\chi}_l
    \begin{pmatrix}
        i q_x \\ k_l
    \end{pmatrix}
    + \tilde{\psi}_l
    \begin{pmatrix}
        k_l \\ -iq_x
    \end{pmatrix}\right]
    +m_2 e^{k_t y} 
    \left[ \tilde{\chi}_t
    \begin{pmatrix}
        i q_x \\ k_t
    \end{pmatrix}
    + \tilde{\psi}_2 
    \begin{pmatrix}
        k_t \\ -iq_x
    \end{pmatrix}\right]
    \right\}
    e^{iq_x x-i\omega t} .
\label{eq: displacement solution}
\end{equation}
The strain tensor at $y=0$ is then
\begin{equation}
\begin{aligned}
    u_{x,x} &=  
        iq_x [m_1 (iq_x \tilde{\chi}_l+k_l \tilde{\psi}_l) + m_2 (i q_x \tilde{\chi}_t + k_t \tilde{\psi}_t)] e^{iq_x x -i\omega t} \\
    u_{y,x} &=
        iq_x [m_1 (k_l \tilde{\chi}_l - i q_x \tilde{\psi}_l) + m_2 (k_t \tilde{\chi}_t - i q_x \tilde{\psi}_t)] e^{iq_x x -i\omega t}  \\
    u_{x,y} &= 
        [k_l m_1 (iq_x \tilde{\chi}_l + k_l \tilde{\psi}_l) + k_t m_2 (iq_x \tilde{\chi}_t + k_t \tilde{\psi}_t)] e^{iq_x x -i\omega t}  \\
    u_{y,y} &= 
        [k_l m_1 (k_l \tilde{\chi}_l - iq_x \tilde{\psi}_l) + k_t m_2 (k_l \tilde{\chi}_t - iq_x \tilde{\psi}_t)] e^{iq_x x -i\omega t} .
\end{aligned}
\label{eq: flat strain}
\end{equation}
The free boundary conditions at $y=0$ are
\begin{equation}
\begin{aligned}
    \sigma_{yx} &= \mu (u_{xy}+u_{yx}) + K_o(u_{yy}-u_{xx}) = 0 ,\\
    \sigma_{yy} &= B(u_{xx}+u_{yy}) + \mu (u_{yy}-u_{xx}) - K_o (u_{xy} +u_{yx})= 0 .
\end{aligned}
\end{equation}
Substituting the strain expression Eq.~\eqref{eq: flat strain} into the boundary conditions, we obtain a linear system
\begin{equation}
\begin{pmatrix}
    BC_{11} & BC_{12} \\
    BC_{21} & BC_{22}
\end{pmatrix}
\begin{pmatrix}
    m_1  e^{iq_x x -i\omega t} \\ m_2 e^{iq_x x -i\omega t}
\end{pmatrix}=0 ,
\end{equation}
where 
\begin{equation}
\begin{aligned}
   BC_{11} &=  [K_o (k_l^2+q_x^2) + 2i\mu q_x k_l]\tilde{\chi}_l+[-2iK_o q_x k_l 
   +\mu (k_l^2 + q_x^2)]\tilde{\psi}_l \\
   BC_{12} &=  [K_o (k_t^2 + q_x^2) + 2i \mu q_x k_t]\tilde{\chi}_t 
   + [-2iK_o q_x k_t  + \mu(k_t^2 + q_x^2) ] \tilde{\psi}_t \\
   BC_{21} &=  [-2i K_o q_x k_l + B(k_l^2 -q_x^2)+\mu (k_l^2 + q_x^2)] \tilde{\chi}_l
   - [K_o(k_l^2+q_x^2)+2i\mu q_x k_l]\tilde{\psi}_l\\
   BC_{22} &= [-2i K_o q_x k_t + B(k_t^2 - q_x^2)+\mu (k_t^2+q_x^2)]\tilde{\chi}_t
   - [K_o (k_t^2 + q_x^2)+2i\mu q_x k_t]\tilde{\psi}_t   . 
\end{aligned}
\end{equation}
A nontrivial solution requires the determinant to vanish
\begin{equation}
\begin{vmatrix}
    BC_{11} & BC_{12} \\
    BC_{21} & BC_{22}
\end{vmatrix} = 0.
\end{equation}
Substituting Eq.~\eqref{eq: kl} and Eq.~\eqref{eq: kt} into this equation yields an equation for the frequency,
\begin{equation}
    R\left(\frac{i \gamma \omega }{\mu q_x^2}, \frac{B}{\mu}, \frac{K_o}{\mu} \right) = 0
\end{equation}
In odd elastic media, the frequency is generally complex, corresponding to wave amplification in one direction and dissipation in the opposite direction. The slope of the complex spectrum can be defined as
\begin{equation}
    S_r = \left| \frac{\text{Im} (i \gamma \omega)}{\text{Re}(i \gamma \omega)} \right|
    = \left| \frac{\text{Im} \lambda }{\text{Re} \lambda} \right|
\label{eq: slope}
\end{equation}
The slope exists for nonzero odd moduli and contrasts with the bulk spectra, which acquires an imaginary component only after the exceptional transition.
\begin{equation}
    S_b = \frac{\text{Im}\left( \sqrt{B^2 - 4K_o^2} \right) }{B+2\mu} 
\end{equation}

\subsection{Spherical caps}
We now generalize the Rayleigh waves to curved systems. Considering a spherical cap, we assume scalar potentials of the form
\begin{equation}
\begin{aligned}
\chi &= \tilde{\chi}  P_l^{m}(\cos \theta) e^{im\phi} e^{-i\omega t},\\
\psi &= \tilde{\psi}  P_l^{m}(\cos \theta) e^{im\phi} e^{-i\omega t}.
\end{aligned}
\end{equation}
where $P(x)$ denotes the Legendre function \cite{haines_spherical_1985, shaqfa_spherical_2021}.
Substituting these solutions into the equations of motion Eq.~\eqref{EOM1}, we obtain the following nontrivial solutions
\begin{equation}
\begin{aligned}
    p(p+1) &= \frac{2 i \gamma \omega r^2}{B+2\mu + \sqrt{B^2-4K_o^2}}  \\
    \begin{pmatrix}
    \Tilde{\chi}_1 \\ \Tilde{\psi}_1
    \end{pmatrix}
    &=
    \begin{pmatrix}
    B + \sqrt{B^2-4K_o^2} \\ 2K_o
    \end{pmatrix}
\end{aligned}
\end{equation}
and 
\begin{equation}
\begin{aligned}
    q(q+1) &= \frac{2 i \gamma \omega r^2}{B+2\mu - \sqrt{B^2-4K_o^2}}  \\
    \begin{pmatrix}
    \Tilde{\chi}_2 \\ \Tilde{\psi}_2
    \end{pmatrix}
    &=
    \begin{pmatrix}
    2K_o \\ B + \sqrt{B^2-4K_o^2}
    \end{pmatrix}
\end{aligned}.
\end{equation}
Here the indices $p$ and $q$ can be complex.
We now combine these two solutions to construct Rayleigh waves:
\begin{equation}
\begin{aligned}
\chi &= \left[m_1 \tilde{\chi}_1  P_p^{m}(\cos \theta) + m_2 \tilde{\chi}_2  P_q^{m}(\cos \theta) \right] e^{im\phi-i\omega t}, \\
\psi &= \left[m_1 \tilde{\psi}_1  P_p^{m}(\cos \theta) + m_2 \tilde{\psi}_2  P_q^{m}(\cos \theta) \right] e^{im\phi-i\omega t}.
\end{aligned}
\end{equation}
The displacement fields follow from the Papkovich--Neuber representation, Eq.~\eqref{Papkovich-Neuber},
\begin{equation}
\mathbf{u} = \left(u_\theta, u_\phi \right)=
\frac{1}{r} \left(
\frac{\partial \chi}{\partial \theta}
+ \frac{1}{\sin\theta}\frac{\partial \psi}{\partial \phi},
\;
\frac{1}{\sin\theta}\frac{\partial \chi}{\partial \phi}
- \frac{\partial \psi}{\partial \theta}
\right),
\label{eq: displacement, spherical}
\end{equation}
where the two components correspond to the $\mathbf{e}_\theta$ and $\mathbf{e}_\phi$ directions on the sphere.
The resulting strain fields are
\begin{equation}
\overline{\nabla} \mathbf{u}=
    \frac{1}{r}
    \begin{pmatrix}
        \displaystyle \frac{\partial u_\theta}{\partial \theta} 
        & \displaystyle \frac{\partial u_\phi}{\partial \theta} \\
        \displaystyle \frac{1}{\sin \theta}\frac{\partial u_\theta}{\partial \phi}-\cot\theta u_\phi 
        & \displaystyle \frac{1}{\sin\theta} \frac{\partial u_\phi}{\partial \phi} + \cot\theta u_\theta 
    \end{pmatrix}.
    \label{eq: spherical strain}
\end{equation}
The free boundary conditions at $\theta=\theta_c$ are
\begin{equation}
\begin{aligned}
    \sigma_{\phi \theta} &= \mu (u_{\theta \phi}+u_{\phi \theta}) + K_o(u_{\phi \phi}-u_{\theta \theta}) = 0 ,\\
    \sigma_{\phi \phi } &= B(u_{\theta\theta}+u_{\phi\phi}) + \mu (u_{\phi\phi}-u_{\theta\theta}) - K_o (u_{\theta \phi} +u_{\phi \theta})= 0 .
\end{aligned}
\end{equation}
In principle, by substituting the strain expression into the boundary conditions, we obtain a linear system and can solve the resulting determinant, analogous to the procedure used for flat systems (Table~\ref{tab:example}). However, the derivatives of the Legendre function (Eq.~\eqref{eq: displacement, spherical} and ~\eqref{eq: spherical strain}) becomes mathematically cumbersome, since the indices $p$ and $q$ are generally complex.

In the limit $\theta_c\to 0$ and $r\to \infty$, while keeping the boundary radius $r\sin\theta_c$ fixed, the spherical geometry reduce to a flat disk. Consequently, the slope of the Rayleigh-wave spectrum in spherical caps can be approximated using the flat-space result in Eq.~\eqref{eq: slope}
\begin{equation}
    S_s \approx S_r.
\end{equation}

\begin{table}[ht!]
\centering
\renewcommand{\arraystretch}{1.3}
\begin{tabular}{
>{\centering\arraybackslash}p{2cm}
>{\centering\arraybackslash}p{2cm}
>{\centering\arraybackslash}p{2cm}
>{\centering\arraybackslash}p{2cm}
}
\hline \hline 
Parameter & Frequency & Wave number & Decay factors  \\ 
\hline
Flat & $\omega$ &  $q_x$ & $k_l$, $k_t$ \\  
Spherical & $\omega$ & $m$ & $p$, $q$ \\  
\hline 
\end{tabular}
\caption{Correspondence of parameters between flat and spherical systems.}
\label{tab:example}
\end{table}

\end{document}